# High-temperature thermoelectric properties of half-Heusler phases $Er_{1-x}Ho_xNiSb$

K. Ciesielski*, K. Synoradzki, I. Wolańska, P. Stuglik, D. Kaczorowski

*Institute of Low Temperature and Structure Research, Polish Academy of Sciences, POB 1410, 50-950 Wrocław, Poland*

**Abstract**

Polycrystalline samples of $Er_{1-x}Ho_xNiSb$ (x = 0, 0.2, 0.3, 0.5, 0.7, 0.8, 1) were characterized by means of x-ray powder diffraction (XRD), scanning electron microscopy (SEM), and optical metallography. The results proved the formation of half-Heusler alloys in the entire composition range. Their electrical transport properties (resistivity, thermoelectric power) were studied in the temperature interval 350-1000 K. The measured electrical resistivity spanned between 5 and 25 µΩm. The maximum thermopower of 50-65 µV/K was observed at temperatures 500-650 K. Replacing Ho for Er resulted in a non-monotonous variation of the thermoelectric power factor (PF = $S^2/\rho$). The largest PF of 4.6 µWcmK$^{-2}$ was found at 660 K for $Er_{0.5}Ho_{0.5}NiSb$. This value is distinctly larger than PF determined for the terminal phases ErNiSb and HoNiSb.

## 1. Introduction

Thermoelectric materials attract much attention of material scientists due to a variety of potential applications, from precision cooling through waste heat recovery, up to autonomous power suppliers without moving parts. Half-Heusler phases have been recognized as prospective thermoelectric materials for moderate-temperature range. So far, the main research focus has been put on *d*-electron compounds with general formula $T^{IV}$(Ni, Co)(Sb, Sn), where $T^{IV}$ is an element from the fourth group of the Periodic Table [1]. Achievements of high figures of merit *ZT* and low production cost have made these materials attractive for large-scale applications [2]. However, some issues involving relatively high thermal conductivity remain unsolved [2]. Recently, exploratory reports have shown that rare-earth (*R*) bearing half-Heusler phases $RT^xSb$ can also exhibit good thermoelectric performance, characterized by fairly low thermal conductivity [3-16]. Based on these predictions, we examined in this work a series of $Er_{1-x}Ho_xNiSb$ samples with the main aim at determining the effect of rare-earth substitution on the magnitude of their thermoelectric power factor.

---

\* Corresponding author. Tel.: +48-71-3954-252; fax: +48-71-3441-029.

*E-mail address:* k.ciesielski@int.pan.wroc.pl

## 2. Experimental

Polycrystalline specimens from the $Er_{1-x}Ho_xNiSb$ series (x = 0, 0.2, 0.3, 0.5, 0.7, 0.8, 1) were synthesized by single-step arc melting performed under argon atmosphere. Purity of starting materials were 99.9 % for erbium and holmium, 99.99 % for nickel, and 99.999 % for antimony. Due to high vapor pressure of antimony additional amount (6 wt.%) of Sb was added beforehand with respect to the ideal composition. The specimens were re-melted and flipped over five times to promote homogeneity.

The products were examined by powder X-ray powder diffraction (XRD) using an Xpert Pro PANanalytical diffractometer. Rietveld analysis was carried out with FullProf software [17]. Density was checked by weighting samples and measuring their dimensions with calliper. Element distribution and topography of the samples were analyzed with a FE-SEM microscope (FEI NovaNanoSEM 230) equipped with an EDS analyzer (EDAX Genesis XM4). Electrical resistivity ($\rho$) and Seebeck coefficient ($S$) measurements were performed in temperature range 350-1000 K under pure helium atmosphere using a Linseis LSR-3 equipment. Experimental uncertainty in these measurements was 3 % for $\rho$ and 5 % for $S$.

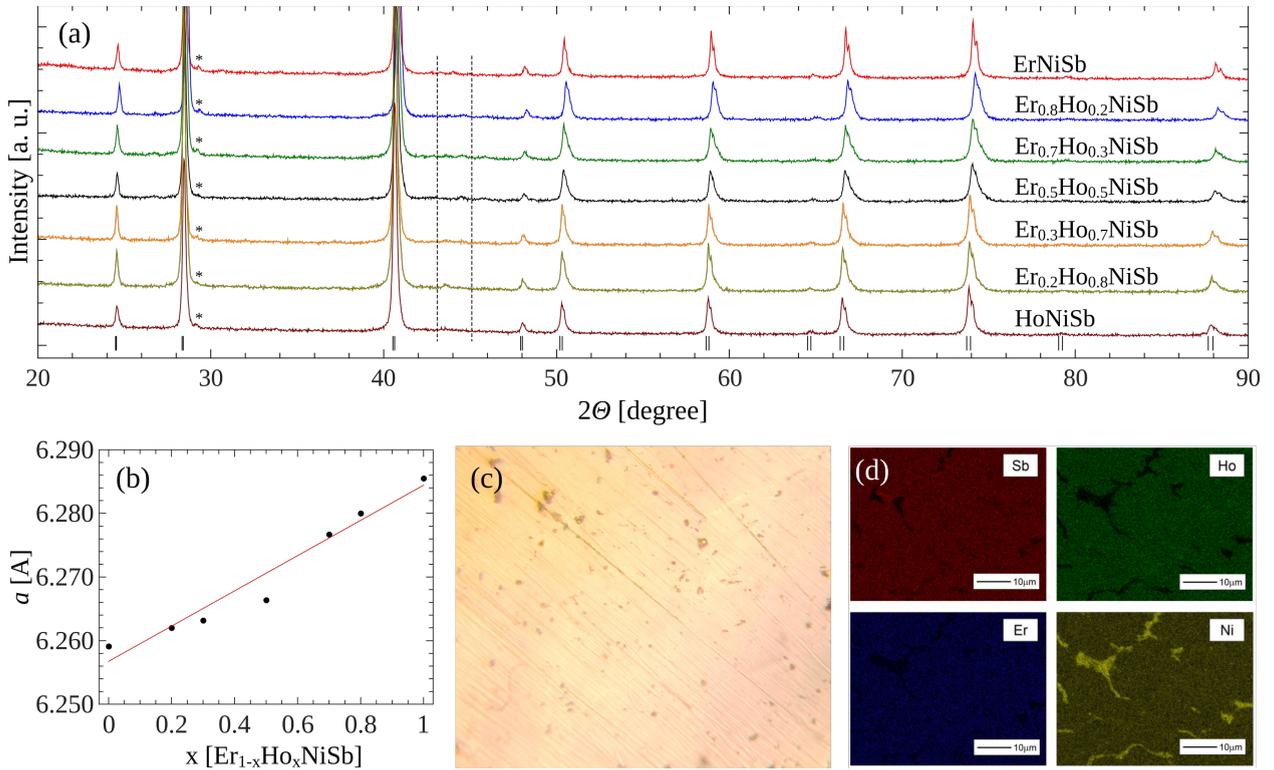

Fig. 1 (a) Powder X-ray diffraction patterns of the $Er_{1-x}Ho_xNiSb$ (x = 0, 0.2, 0.3, 0.5, 0.7, 0.8, 1) samples. Ticks on the bottom mark the positions of the Bragg reflections due to the half-Heusler main phase. Impurity phases $Er_2O_3/Ho_2O_3$ and $Ni_{1-x}Sb_x$ are marked by asterisks and vertical dashed lines, respectively. (b) Lattice parameter versus Ho-content in the $Er_{1-x}Ho_xNiSb$ samples. Red straight line demonstrates conservation of the Vegard's law. (c) Optical metallographic microscope image (x500) of $Er_{0.2}Ho_{0.8}NiSb$. Dark regions are due to cavities. (d) SEM EDS chemical composition maps for $Er_{0.2}Ho_{0.8}NiSb$.

## 3. Results and discussion

Fig. 1a presents the powder XRD patterns of the $Er_{1-x}Ho_xNiSb$ samples. All the obtained samples were found to crystallize with the cubic MgAgAs-type crystal structure, characteristic of the half-Heusler phases. Some small spurious Bragg peaks are visible at around 29 deg and in the region 43 – 45 deg, which can be attributed to $Er_2O_3/Ho_2O_3$ and $Ni_{1-y}Sb_y$ impurity phases, respectively. The cubic lattice parameter derived for ErNiSb is close to those reported in the literature [13-15, 18-20], while that obtained for HoNiSb is slightly larger (by about 0.5 %) than the literature values [13, 20-21]. In between the terminal compositions, the lattice parameter obeys the Vegard's law (see Fig. 1b) that proves full mutual solubility of the two compounds. Fig. 1c shows an exemplary optical metallography image of the alloy $Er_{0.2}Ho_{0.8}NiSb$. The material appears on this photograph fairly homogeneous, with small darker regions due to cavities, which were found unavoidable in arc-melting synthesis of the Sb-containing samples. The presence of such cavities limited precise determination of the geometrical factor of the specimens used in electrical transport measurements (see below), and led to their high mechanical fragility. Representative SEM-EDS element distribution maps obtained for the same alloy are given in Fig. 1d. The EDS data proved a uniform distribution of erbium and holmium in the samples studied. Nonetheless, the microprobe analysis indicated the presence of parasitic $Ni_{1-y}Sb_y$ solid solution with the index y = 0.1-0.5. Formation of this secondary phase resulted in the appearance of vacancies on the Ni-atom sites in the half-Heusler matrix in the amount of up to 20%, as resolved with the Rietveld refinements (see Tab. 1). Here, it is worth recalling that similar finding of Ni-atom vacancies up to 15% was reported for single crystals of ScNiSb [22] and polycrystalline samples of LuNiSb [23]. The amount of the impurity phases was too small to be precisely obtained through Rietveld analysis, however it was estimated to be less than several percent.

The temperature dependencies of the electrical resistivity of the $Er_{1-x}Ho_xNiSb$ samples are displayed in Fig. 2a. For ErNiSb, the resistivity magnitude near room temperature is smaller than the values reported in the literature (29 μΩm [13] and 52 μΩm [14]). The discrepancy may be attributed to some differences in microstructure or stoichiometry of the specimens investigated. As apparent from the figure, the $\rho(T)$ curves can be well reproduced with a simple formula $\rho = 1/(\sigma_s + \sigma_m)$ that assumes two independent channels of charge transfer: semiconducting-

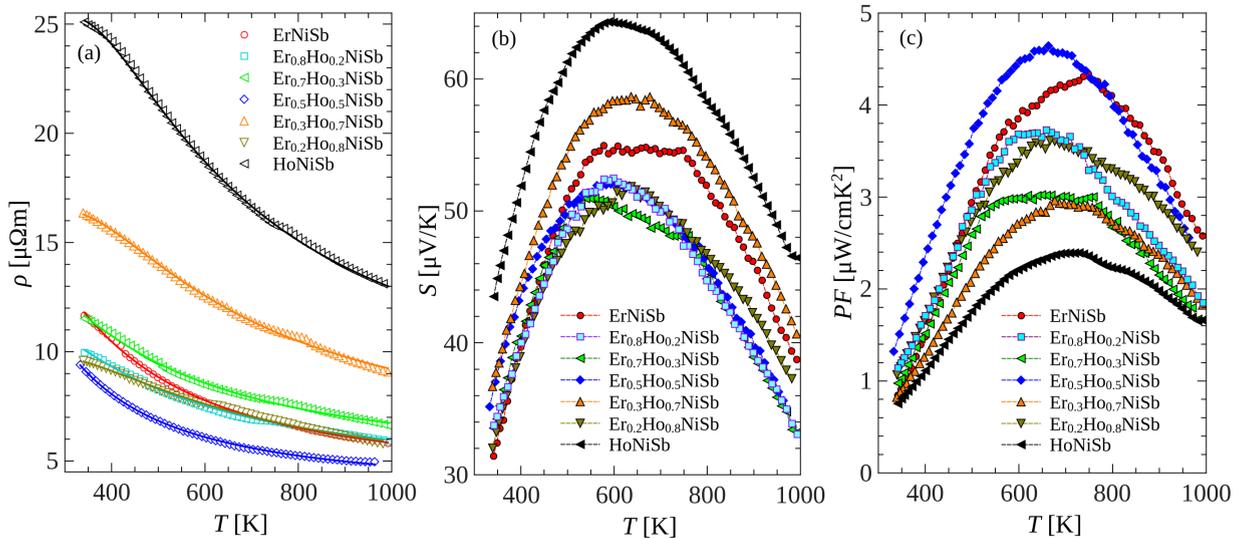

Fig. 2 (a) Temperature variations of the electrical resistivity of the $Er_{1-x}Ho_xNiSb$ (x = 0, 0.2, 0.3, 0.5, 0.7, 0.8, 1) alloys. Solid lines represent the modelling described in the text. (b) Temperature dependencies of the thermoelectric power of the $Er_{1-x}Ho_xNiSb$ (x = 0, 0.2, 0.3, 0.5, 0.7, 0.8, 1) alloys. Dashed lines serve as a guide to the eye. (c) Temperature variations of the power factor calculated for the $Er_{1-x}Ho_xNiSb$ (x = 0, 0.2, 0.3, 0.5, 0.7, 0.8, 1) alloys. Dashed lines serve as a guide to the eye.

Table 1. Density and relative density with respect to theoretical values calculated from lattice parameter (in brackets), occupation of 4c Wyckoff site obtained through Rietveld analysis, and parameters characterizing the electrical conductivity in the $Er_{1-x}Ho_xNiSb$ (x = 0, 0.2, 0.3, 0.5, 0.7, 0.8, 1) samples.

|  | Density [g/cm3] | 4c (Ni) occ. (%) | $\rho_0$ ($\mu\Omega$m) | $a$ ($\mu\Omega$m/K) | $\sigma_0$ (1/$\mu\Omega$m) | $E_g$ (meV) |
|---|---|---|---|---|---|---|
| ErNiSb | 8.32 (88 %) | 96 | 1.2 | 0.097 | 0.281 | 95 |
| $Er_{0.8}Ho_{0.2}NiSb$ | 7.83 (83 %) | 83 | 11.0 | 0.010 | 0.247 | 124 |
| $Er_{0.7}Ho_{0.3}NiSb$ | 8.35 (89 %) | 80 | 13.7 | 0.007 | 0.211 | 129 |
| $Er_{0.5}Ho_{0.5}NiSb$ | 8.15 (87 %) | 80 | 20.6 | 0.011 | 0.292 | 86 |
| $Er_{0.3}Ho_{0.7}NiSb$ | 7.77 (84 %) | 95 | 3.6 | 0.058 | 0.209 | 140 |
| $Er_{0.2}Ho_{0.8}NiSb$ | 8.05 (87 %) | 86 | 6.3 | 0.018 | 0.302 | 146 |
| HoNiSb | 7.84 (85 %) | 92 | 1.4 | 0.103 | 0.164 | 152 |

and metallic-like, with electrical conductivities $\sigma_s$ and $\sigma_m$, respectively. The two terms can be expressed as $\sigma_m = (\rho_0 + aT)^{-1}$ and $\sigma_s = \sigma_0 \exp(-E_g/2k_BT)$, where $\rho_0$ is the residual resistivity, $aT$ accounts for scattering on phonons, and $E_g$ is the energy gap. While the semiconducting channel is commonly considered to be intrinsic for stoichiometric half-Heusler $RT^xSb$ phases, the metallic conductivity may be assigned to a narrow impurity band located within the energy gap of the matrix. The appearance of such in-gap metallic states can be attributed to structural disorder that is inherent to most half-Heusler compounds. Applying the proposed charge transfer model to the experimental resistivity data of the alloys $Er_{1-x}Ho_xNiSb$ yielded the parameters listed in Table 1. The energy gap increases with increasing the holmium content except for $Er_{0.5}Ho_{0.5}NiSb$ that exhibits the smallest $E_g$ value within the series. Microscopic origin of this anomaly is unclear. It is worth noting that the so-derived values of the energy gap in the terminal phases ErNiSb and HoNiSb are of similar magnitude to those reported in the literature ($E_g$ = 80 and 90 meV, respectively [13]).

As can be inferred from Fig. 2a, the gradual substitution of Ho for Er in the $Er_{1-x}Ho_xNiSb$ alloys does not result in any systematic change in $\rho(T)$. Modification of electrical resistivity values cannot also be correlated with the sample densities (see Tab. 1). Remarkably, the electrical resistivity of the $Er_{0.5}Ho_{0.5}NiSb$ sample is smaller than the resistivity of the terminal compounds. This finding can be attributed to a particular microstructure of the measured specimen, which probably affects the charge transport in a more distinct manner than changes occurring in the electronic structure of the alloy series studied. Non-monotonic change in the electrical resistivity upon substitutions on rare-earth-equivalent site has already been reported for some half-Heusler phases, e.g. for $Hf_{1-x}Zr_xNiSn_{0.98}Sn_{0.2}$ [24]. Furthermore, for the solid solution ZrNiSn-ErNiSb [13], the alloy $Zr_{0.75}Er_{0.25}NiSn_{0.75}Sb_{0.25}$ was found to exhibit an order of magnitude larger electrical conductivity than the parent ternary compounds.

Fig. 2b shows the temperature dependencies of the Seebeck coefficient measured for the $Er_{1-x}Ho_xNiSb$ alloys. For each sample, the thermopower is positive in the whole temperature range investigated, which suggests that the charge transport in these materials is dominated by holes. Overall shape of the $S(T)$ curves is characteristic for *p*-type materials with wide maxima occurring within the temperature interval 500-650 K. The thermopower magnitude is rather moderate, up to 64 µV/K in 590 K for HoNiSb. As seen in Fig. 2b, the substitution of Er for Ho has a rather minor effect on the Seebeck coefficient. It should be recalled that the thermopower data available in the literature for the terminal compounds are highly dispersed. The room temperature $S$ values range for HoNiSb from -19 µV/K [25] to 38 µV/K [13], while the values reported for ErNiSb are 30 µV/K [15], 160 µV/K [13] and 248 µV/K [14]. It is worth noting that the result obtained in our study is close to that quoted in Ref. 15.

The temperature variations of the power factor, PF = $S^2/\rho$, calculated for the $Er_{1-x}Ho_xNiSb$ samples are presented in Fig. 2c. The magnitude of PF is rather high, being of the order of several µWcm$^{-1}$K$^{-2}$. The substitution on the rare-earth site brings about an enhancement of PF with respect to that of the terminal phases. The highest PF = 4.64 µWcm$^{-1}$K$^{-2}$ was obtained at 660 K for $Er_{0.5}Ho_{0.5}NiSb$. This value is about 20% larger than PF observed for ErNiSb and nearly twice that determined for HoNiSb.

**4. Summary**

The half-Heusler compounds ErNiSb and HoNiSb were found to exhibit a full mutual solubility. The prepared alloys $Er_{1-x}Ho_xNiSb$ appeared almost single-phase with the cubic lattice parameter obeying the Vegard's law for $0 \leq x \leq 1$. However, sizable nonstoichiometry was detected in a form of Ni-atom deficiency in the main half-Heusler phase (up to 20 at.%). The electrical transport in the $Er_{1-x}Ho_xNiSb$ samples was described by a model comprising parallel semiconducting- and metallic-like charge transfer channels. The substitution of Ho for Er was found to moderately affect the thermoelectric properties of the alloys. Nevertheless, the power factor determined for $Er_{0.5}Ho_{0.5}NiSb$ was larger by about 20% (at 660 K) than its magnitude for ErNiSb. In the next step of the research, one should check if the intentionally introduced disorder on the rare-earth atom site brings about sizable reduction in the heat conductivity of the $Er_{1-x}Ho_xNiSb$ alloys and hence significant improvement in the thermoelectric performance of the system studied.

**Acknowledgements**

This work was supported by the National Science Centre (Poland) under research grant no. 2015/18/A/ST3/00057.